\documentclass[10pt,journal,compsoc]{IEEEtran}
\usepackage{amssymb}
\ifCLASSINFOpdf
   \usepackage[pdftex]{graphicx}
\else
   \usepackage[dvips]{graphicx}
\fi
\usepackage{subfigure}

\usepackage{float}
\usepackage{setspace}
\usepackage[cmex10]{amsmath}
\usepackage{mdwmath}
\usepackage{setspace}
\usepackage{multirow}
\usepackage{array}
\usepackage{cases,multicol}
\usepackage[lined,boxed,commentsnumbered, ruled]{algorithm2e} 
\usepackage{amsthm} 
\renewenvironment{proof}{{\bf \emph{Proof}: }}{\hfill $\blacksquare$}
\usepackage{chemarrow}
\usepackage{extarrows}
\usepackage{color}
\usepackage{epstopdf}
\allowdisplaybreaks[4]
\usepackage[colorlinks,
            linkcolor=black,
            anchorcolor=black,
            urlcolor=black,
            citecolor=black]{hyperref}
\usepackage[hyphenbreaks]{breakurl}
\newtheorem{theorem}{Theorem}

\newtheorem{lemma}{Lemma}
\newtheorem{definition}{Definition}

\newcommand{\methodname}{{\tt{FAMuS}}}

\begin{document}
\title{Fairness-Aware Multi-Server Federated Learning Task Delegation over Wireless Networks}

\author{Yulan Gao, \IEEEmembership{~Member, IEEE,}
Chao Ren, \IEEEmembership{~Member, IEEE,}
and Han Yu, \IEEEmembership{~Senior Member, IEEE}
\thanks{ Y. Gao and H. Yu are with the School of Computer Science and Engineering, Nanyang Technological University, 639798 Singapore (e-mail: yulan.gao@ntu.edu.sg, han.yu@ntu.edu.sg).}
\thanks{C. Ren is with the School of Electrical and Electronic Engineering, Nanyang Technological University, 639798 Singapore, and the School of Electrical Engineering and Computer Science, KTH Royal Institute of Technology, Sweden. (e-mail: chao.ren@ntu.edu.sg).}
}

\markboth{~}
{Shell \MakeLowercase{\textit{et al.}}: }

\IEEEtitleabstractindextext{
\begin{abstract}
In the rapidly advancing field of federated learning (FL), ensuring efficient FL task delegation while incentivizing FL client participation poses significant challenges, especially in wireless networks where FL participants' coverage is limited.
Existing Contract Theory-based methods are designed under the assumption that there is only one FL server in the system (i.e., the monopoly market assumption), which in unrealistic in practice.
To address this limitation, we propose \underline{F}airness-\underline{A}ware \underline{Mu}lti-\underline{S}erver FL task delegation approach (\methodname{}), a novel framework based on Contract Theory and Lyapunov optimization to jointly address these intricate issues facing wireless multi-server FL networks (WMSFLN).
Within a given WMSFLN, a task requester products multiple FL tasks and delegate them to FL servers which coordinate the training processes.
To ensure fair treatment of FL servers \cite{Shi-et-al:2023FAFL}, \methodname{}
establishes virtual queues to track their previous access to FL tasks, updating them in relation to the resulting FL model performance.
The objective is to minimize the time-averaged cost in a WMSFLN, while ensuring all queues remain stable.
This is particularly challenging given the incomplete information regarding FL clients' participation cost and the unpredictable nature of the WMSFLN state, which depends on the locations of the mobile clients.
Extensive experiments comparing \methodname{} against five state-of-the-art approaches based on two real-world datasets demonstrate that it achieves 6.91\% higher test accuracy, 27.34\% lower cost, and 0.63\% higher fairness on average than the best-performing baseline.
\end{abstract}

}

\maketitle
\IEEEdisplaynontitleabstractindextext

\IEEEpeerreviewmaketitle

\section{Introduction}
In the evolving landscape of machine learning (ML), centralized learning approaches have traditionally taken the center stage \cite{fourati2021survey}.
As data generation scales up and concerns about data privacy become more widespread, such an approach faces inherent challenges \cite{voigt2017eu, kairouz2021advances}.
A promising way out of this gridlock is federated learning (FL) \cite{Lyu-et-al:2022}, which is a decentralized learning paradigm where devices, from smartphones to industrial IoT sensors, perform localized model training and collaboratively build global ML models.
Rather than transmitting raw data, only model updates are uploaded to an FL server for coordination and aggregation, thereby achieving enhanced privacy preservation and the ability to utilize diverse, real-world data sources \cite{yang2019federated}.
This salient feature of FL is further empowered by the development of wireless networks \cite{niknam2020federated,liu2020federated}.
Today, exploring FL in the context of wireless networks emerges as an important field of research in areas such as connectivity, scalability, real-time collaboration, and energy efficiency.

Navigating the complexities of optimizing FL for practical applications, we are presented with challenges that span both technical and economical dimensions.
Among the early contributions in the technical dimension, \cite{tran2019federated, pang2022incentive} focused on proving the balance between computation and communication is crucial theoretically and experimentally.
Since FL requires constant communication between FL server and different mobile clients to continuously refine the model until convergence is achieved, it induces complex dynamics: the pursuit of higher local accuracy requires a large number of clients committing more computational resources \cite{ma2017distributed}.
Delving deeper, however, and the economic dimension becomes an equally compelling problem. The iterative nature of FL requires the ongoing engagement of numerous clients who need to be motivated.
Furthermore, it is unrealistic to expect clients to consistently participate throughout the FL training process, especially considering the inherent mobility of FL client devices and their Thus, incentives need to be provided to motivate FL client participation.

A typically FL incentive mechanism needs to perform three distinct tasks \cite{zeng2021comprehensive}: 1) FL task delegation, 2) FL client contribution assessment, and 3) allocation of rewards.
One of established theoretical tools for incentive mechanism design is the Contract Theory \cite{dai2018contract}.
Contract theory focuses on the formulation of optimal agreements among parties with distinct interests and varying degrees of information.
In FL, two main lines of research for Contract Theory-based incentivization can be identified depending on the assumptions: 1) asymmetric contracts, and 2) multi-dimensional contracts. In the first branch, \cite{lim2020hierarchical,kang2019incentivec,kang2019incentive} addressed the information asymmetry between the FL server and FL clients. They link resources to rewards via contract lists, focusing on local computing power as a decision variable, and operating under a scenario with limited knowledge about a client's type.
The second branch delves into the provisioning of multi-dimensional resources in FL.
In \cite{ding2020incentive,lu2021toward}, the authors formulated contract design using dataset size and communication time as key parameters. They suggest that
adhering to communication constraints can simplify the incentive mechanism, thereby deriving optimal rewards and establishing benchmarks for dataset size and maximum communication duration.

Existing Contract Theory-based methods are designed under the assumption that there is only one FL server in the system (i.e., the monopoly market assumption). This assumption is unrealistic in practice.
In addition, they also do not adequately address the challenge facing FL in wireless networks in which mobile devices (i.e., FL clients) have limited communication coverage, forcing FL server to only work with clients it can directly reach.

To bridge these important gaps, we propose \underline{F}airness-\underline{A}ware \underline{Mu}lti-\underline{S}erver FL task delegation approach (\methodname{}).
It draws upon Contract Theory and Lyapunov optimization to manage task delegation in the context of providing incentives to FL clients in a wireless multi-server FL network (WMSFLN) setting.
Under our problem setting, multiple FL servers coexist with a task requester (TR) which is responsible for delegating FL task requests to the FL servers.
The TR's service area is segmented into clusters, each managed by one FL server.
Since clients can move freely throughout the entire area, the candidate client pool for each FL server changes dynamically.
Thus, \methodname{} is designed to balance the performance and fairness during task delegation.
Mobile clients are categorized based on their associated participation costs.
FL servers offer contracts comprising various items that control the participation decision and rewards for specific client types, with the goal of eliciting truthful type information from clients while minimizing time-averaged WMSFLN cost.
Following Lyapunov optimization, \methodname{} can deal with the unknown distribution of clients and the time-coupled queue stability constraints. For enhanced scalability, the task delegation and participants are controlled independently by each FL server based only on local information.

We conduct extensive experiments on two real-world datasets to assess and compare the proposed \methodname{} against five state-of-the-art approaches. Our findings indicate that \methodname{} outperforms the best baseline by an average 6.91\% and 0.63\% improvement in test accuracy and fairness, respectively, and a 27.34\% reduction in costs.

\section{Related Work}\label{related}
Current incentive mechanisms in FL primarily on Game Theory \cite{weng2021fedserving}, auctions \cite{zeng2020fmore}, blockchains \cite{bao2019flchain} and Contract Theory. Among them, Contract Theory-based methods are the most closely related to our work.
These techniques are mainly employed by node selection ({\em a.k.a.} client selection) and payment allocation. Among the contributions in the area of game theory-based incentive mechanisms, \cite{pandey2020crowdsourcing,zeng2020fmore} developed the Stackelberg game-based incentive framework to optimize clients recruitment for utility maximization during training. In the spirit of these works, Weng \textit{et. al.} \cite{weng2021fedserving} focused on designing incentive mechanism based on Bayesian game theory, emphasizing privacy protection, truthfulness, and accuracy. Lim \textit{et. al.} \cite{lim2021decentralized} introduced a framework combining resource allocation and incentive mechanism in hierarchical FL, utilizing evolutionary game theory and auction. Within the field of auction-based incentive mechanism studies, \cite{le2020auction} proposed a randomized auction framework to incentive clients in wireless network scenario. Zeng \textit{et. al.} \cite{zeng2020fmore} developed FMore, a multi-dimensional incentive method utilizing procurement auction for FL in the context of mobile edge computing (MEC).Similarly, a quality-aware auction scheme in multi-task learning scenario was proposed in \cite{deng2021fair}. \cite{jiao2020toward} optimized the social welfare of wireless FL system by a reverse multi-dimensional auction (RMA) mechanism, employing a randomized and greedy approach for participants selection. FL's security and privacy can be enhanced by  incorporating blockchain into incentive mechanisms, leveraging its inherent robustness and tamper-resistance features. Bao \textit{et. al.} \cite{bao2019flchain} proposed FLChain, building a transparent, trust, and incentive FL ecosystem. Similarly, Toyoda \textit{et. al.} \cite{toyoda2019mechanism} used contest theory to model and analyze an incentive-aware mechanism for blockchain-assisted FL system. In the following, we briefly summarize the related work of contract theory-based incentive mechanisms.
Contract Theory focuses on the formulation of optimal agreements among parties with distinct interests and varying degrees of information.
However, there are only a limited number of Contract Theory-based studies on FL.
Authors in \cite{kang2019incentivec} adopted Contract Theory to motivate high-quality FL clients to perform model updates, thereby eliminating unreliable ones.
In addition, they extended this approach by integrating Contract Theory with reputation modeling and blockchain \cite{kang2019incentive} to help FL servers gain deeper understanding about clients' behaviour patterns.
A notable study that provides a theoretical analysis of multi-dimensional incentive mechanism design in FL, considering three distinct levels of information asymmetry can be found in \cite{ding2020incentive}.
A multi-dimensional Contract Theory-based reliable incentive mechanism was designed for the UAV-aided Internet-of-Vehicles scenario \cite{lim2021towards}. Lu \textit{et. al.} \cite{lu2021toward} focused on fairness-aware, time-sensitive task allocation in asynchronous FL and developed a multi-dimensional Contract-Theoretic approach to optimize FL model accuracy.

Different from these approaches, \methodname{} does not rely on the monopoly market assumption. It is designed to achieve fairness-aware task delegation in WMSFLN, combining Lyapunov optimization with Contract Theory to address the challenges of incomplete information about the FL clients' costs and their unknown distributions.

\section{Preliminaries}
\subsection{WMSFLN System Model}
\begin{figure}[!t]
	\centering
	\includegraphics[width=1\linewidth]{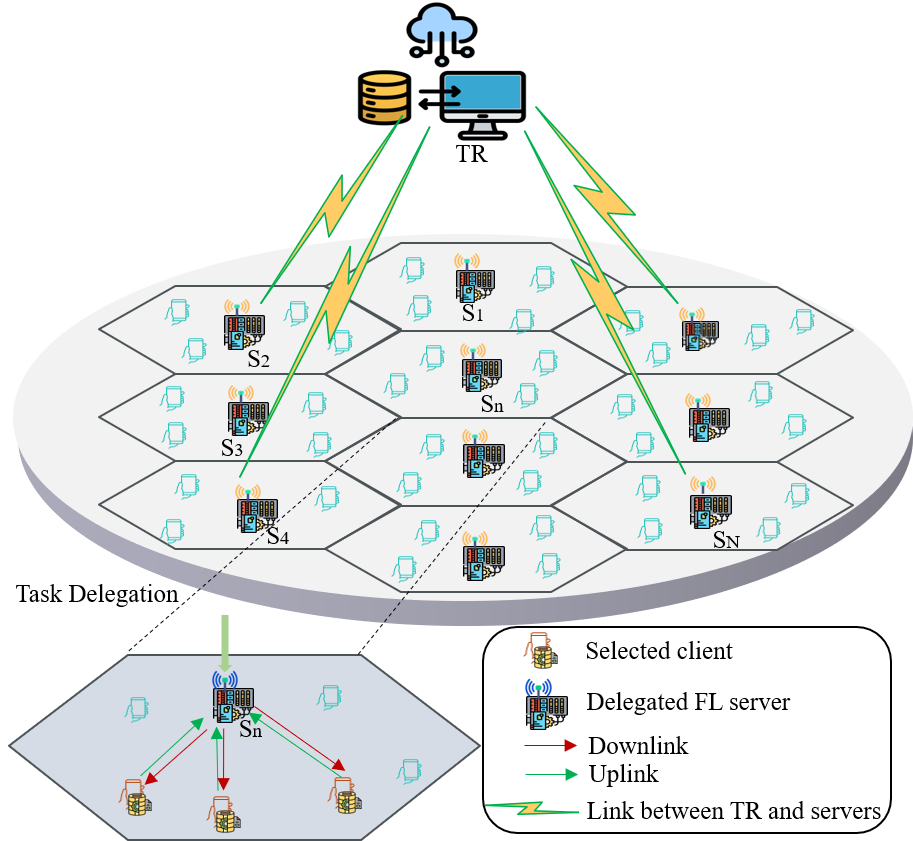}
	\caption{A WMSFLN consists of a task requester and multiple FL servers. Each FL server coordinates the FL training process involving multiple mobile FL clients within its cluster. Clients can move from cluster to cluster.}
	\label{fig:1}
\end{figure}

Consider a WMSFLN as depicted in Figure \ref{fig:1}, wherein the TR receives FL tasks from users and delegates them to different FL servers.
The FL servers then coordinate the training of these tasks by engaging suitable FL clients.

Let ${\mathcal S}=\{s_1, s_2, \ldots, s_N\}$ be the set of FL servers, with their index set denoted as ${\mathcal N}=\{1,\ldots, N\}$.
We assume that a WMSFLN comprises $N$ clusters, each managed by a dedicated FL server.
Let ${\mathbf R}_0\subseteq {\mathbb R}^2$ represent the entire area covered by one WMSFLN, with the associated area covered by $s_n$ denoted as ${\mathbf R}_n$\footnote{Due to the movements of clients into and out of clusters, the coverage of each cluster is dynamic (i.e., $|{\mathbf R}_n|\neq |{\mathbf R}_{n'}|, \forall n'\in{\mathcal N}\setminus \{n\}$, where $|\cdot|$ represents the set cardinality).}.
In addition, within the WMSFLN ${\mathbf R}_0$, there exist a set of $M$ independently moving clients, denoted as ${\mathcal M}=\{1,2, \ldots, M \}$ and labeled as $c{_1}, c_{2}, \ldots, c_{M}$.
For simplicity, we assume all servers and their associated clients initiate a new global training round simultaneously, with each server broadcasting global model parameters to selected clients and awaiting their updates. To maintain synchronicity across servers performing diverse tasks, mechanisms like a central coordinator or consensus algorithm can be employed.

Building on \cite{ding2020incentive}, we adopt the typical FL with a one-step local update.
Without loss in generality, we assume that the TR releases tasks at intervals of $\tau$.
Each global round is constrained by a duration of $\triangle_t$.
Here, $t\in{\mathbb N}_0$ acts as an index, marking each individual time slot.
Consequently, this setup results in the WMSFLN operating on a structured and time-slotted basis with the time axis uniformly divided into segments of equal duration.
Given that a time slot $\triangle_t$ is small, all network parameters (e.g., locations and data quality of clients, as well as channel quality) are treated as stable within a given time slot.
\subsection{Task Delegation and Client Selection}
We first outline the joint task delegation and client selection problem within the WMSFLN.
Define ${\mathcal C}_t^n=\{m\in{\mathcal M}| ~\text{loc}_t^m\in{\mathbf R}_n\}, n\in{\mathcal N}$, as the subset of clients located in a cluster ${\mathbf R}_n$ in time slot $t$ and satisfying $\cap_{n\in{\mathcal N}} {\mathcal C}_t^n=\emptyset$ and $\cup_{n\in{\mathcal N}}{\mathcal C}_t^n={\mathcal M}$, where $\text{loc}_t^m$ denotes the location of client $m$ in slot $t$.
In every slot $t$, satisfying $\sum_{i=0}^{t-1}\triangle_i \mod \tau=0$, the TR releases $|{\mathcal J}_t|$ FL tasks. These tasks are denoted by the set ${\mathcal J}_t=\{J_t^1, \ldots, J_t^{|{\mathcal J}_t|}\}$, with the constraint $|{\mathcal J}_t|\leq N$.
In addition, the set of indices for the tasks is denoted as ${\mathcal K}=\{1, 2, \ldots, K\},$ with $K\leq N.$

Let ${\mathbf A}_{\mathbf{loc}_t}=\{{\mathbf a}_t\}$ be the task delegation action space, where ${\mathbf a}_t=\{a_t^{k,n}\in\{0, 1\}|k\in{\mathcal K}, n\in{\mathcal N}\}$ denotes the task delegation matrix.
In this matrix,  $a_t^{k,n}=1$, if task $k$ is delegated to server $s_n$ in slot $t$; otherwise, $a_t^{k,n}=0.$
It is noted that, the finite set ${\mathbf A}_{\mathbf{loc}_t}$ is defined according to the current WMSFLN network state $\mathbf{loc}_t$, where $\mathbf{loc}_t=\{\text{loc}_t^m~|~m\in{\mathcal M}\}$ and $\text{loc}_t^m$ represents the location of client $m$ at time $t$.
The task delegation action space is constrained by:
\begin{subequations}
\begin{align}
\sum\nolimits_{n\in{\mathcal N}}a_t^{k,n}&\leq 1, \forall k\in{\mathcal K}, \label{eq:3a}\\
\sum\nolimits_{k\in{\mathcal K}}a_t^{k,n}&\leq 1, \forall n\in{\mathcal N}, \label{eq:3b}
\end{align}
\end{subequations}
where Eq. \eqref{eq:3a} and Eq. \eqref{eq:3b} indicate that, at time slot $t$, one FL server can coordinate the training of only one FL task, and an FL task can only be assigned to  one server, respectively.

Let ${\mathbf b}_t=\{b_t^{n,m}\in\{0, 1\}|m\in{\mathcal M}, n\in{\mathcal N}\}\in{\mathbf B}_{\mathbf{loc}_t}$ denote the clients selection matrix, such that $b_t^{n,m}=1$, if client $m$ is selected by the FL server $s_n$ at slot $t$. The finite set ${\mathbf B}_{\mathbf{loc}_t}$ of possible values of ${\mathbf b}_t$ is constrained by:
\begin{align}\label{constraint2}
b_t^{n,m}\leq {\mathbf 1}_{[\sum_{k\in{\mathcal K}}a_t^{n,k}\neq 0, m\in{\mathcal C}_t^n]}, \forall m\in{\mathcal M},
\end{align}
where ${\mathbf 1}_{[\text{condition}]}$ is an indicator function. Its value is $1$ iff [condition] is true; otherwise, it evaluates to $0$.
Eq. \eqref{constraint2} ensures that client $m$ can only be selected if it falls within server $n$'s coverage area and the server has pending training tasks to execute.

\subsection{Basics of Federated Learning}
In a typical FL scenario, we consider a data tuple $(x_i, y_i)$, where $x_i$ is the input, and $y_i$ represents the corresponding label.
The aim of the learning process is to identify the optimal model parameter, denoted as ${\pmb\omega}$, capable of accurately forecasting the label $y_i$ using the input $x_i$.
The prediction is denoted as $\tilde{y}(x_i;{\pmb\omega})$.
The difference between this prediction $\tilde{y}(x_i;{\pmb\omega})$ and the true label $y_i$ is measured by the prediction loss function $f_i(\pmb\omega)$.
When a client $m$ utilizes a dataset ${\mathcal D}_m$, encompassing $d_m$ data points for training the model, the client's loss function is ascertained by calculating the mean prediction loss across all data points $i$ included in ${\mathcal D}_m$ as:
\begin{align}\label{eq:1}
F_m(\pmb\omega)=\frac{1}{d_m}\sum\nolimits_{i\in{\mathcal D}_m}f_i(\pmb\omega).
\end{align}
The optimal model parameter, denoted as ${\pmb\omega}^{*}$, aims to minimize the global loss function, formulated as a weighted average of the loss functions across all clients:
\begin{align}\label{eq:2}
{\pmb\omega}^{*}=\arg\min_{\pmb\omega}f(\pmb\omega)=\arg\min_{\pmb\omega}\sum\nolimits_{m=1}^M\frac{d_m}{d}F_m(\pmb\omega),
\end{align}
where $d=\sum_{m=1}^Md_m$ is the total data size of all clients.

\section{The Proposed \methodname{} Approach}

\subsection{FL Clients' Payoffs and WMSFLN Cost}
We let $d_t^m$ denote the required training dataset size of client $c_m$ in slot $t$.
For each client, the cost consists of the local model uploading cost and the local model training cost. It is reasonable to assume that the cost for model transmission is proportional to the data rate from clients to the corresponding FL servers, and the cost for local model training is proportional to the local dataset size \cite{tran2019federated}.
The cost incurred by client $c_m$ under FL server $s_n$ at slot $t$, $c_t^{n,m}$, is expressed as:
\begin{align}\label{eq:4}
c_t^{n,m}=\alpha^mR_t^{n,m}+\beta^md_t^m, \forall n\in{\mathcal N}, m\in{\mathcal M},
\end{align}
where $\alpha^m>0$ is the cost incurred by $c_m$ per unit data rate. $\beta^m$ is the per unit data-usage cost. $R_t^{n,m}$ is the instantaneous data rate of transmission link from client $c_m$ to server $s_n$ at slot $t$. Based on Shannon Formula,  $R_t^{n,m}$ is expressed as:
\begin{align}\label{eq:5}
R_t^{n,m}=B_t^{n,m}\log_2\left(1+\frac{p_t^mG_t^{n,m}}
{N_0B_t^{n,m}}\right),
\end{align}
where $G_t^{n,m}$ and $p_t^m$ represent the channel gains from client $m$ to server $n$ and the transmit power of client $m$ at slot $t$, respectively. $N_0$ is the Gaussian white noise power spectral density. The term $B_t^{n,m}$ represents the bandwidth allocated for uploading model parameters from client $c_m$ to the FL server $s_n$.
To be fair, each participating client in cluster ${\mathbf R}_n$ is allocated an equal share of the communication bandwidth, denoted as $B_t^{n,m}={B_n}/{\sum_{m'\in{\mathcal C}_t^n}b_t^{n,m'}}$, where $B_n$ is the total bandwidth of the assigned licensed band.

It is worth noting that both data rate $R_t^{n,m}$ and data size $d_t^m$ are only observable by client $c_m$.
As a result, only $c_m$ is privy to its participation cost $c_t^{n,m}$ for a training task.
While FL servers can obtain this information, doing so requires additional control signaling, potentially leading to significant delays and increased cost for client $c_m$.
Next, we define the type of client $c_m$ participating in the training task under FL server $s_n$ as $\gamma_t^{n,m}=1/c_t^{n,m}$, and $\gamma_t^{n,m}>0.$
The type $\gamma_t^{n,m}$ contains private information about the cost incurred by client $c_m$ when participating in training task associated with FL server $s_n$.
Consequently, lower values of $\gamma_t^{n,m}$ indicate a higher participation expense.
Let ${\mathbf r}_t=\{r_t^{n,m}\geq 0 | m\in{\mathcal M}, n\in{\mathcal N}\}$ denote the clients' reward matrix, where $r_t^{n,m}$ is the reward of client $c_m$ received from participating in tasks coordinated by FL server $s_n$.
The payoff for client $c_m$ from participating in a training task under FL server $s_n$ is:
\begin{align}\label{eq:7}
u_t^{n,m}=r_t^{n,m}-\frac{b_t^{n,m}}{\gamma_t^{n,m}}.
\end{align}

In addition, in the WMSFLN system, the total cost for the TR include the expected accuracy loss of the FL servers, service fees to these servers, and rewards given to participating clients.
Hence, the total WMSFLN system cost at slot $t$ is expressed as follows:
\begin{equation}\label{eq:8}
c({\mathbf a}_t,{\mathbf b}_t, {\mathbf r}_t, \mathbf{loc}_t)=\sum\nolimits_{n\in{\mathcal N}}c^n({\mathbf a}_t^n,{\mathbf b}_t^n, {\mathbf r}_t^n, \mathbf{loc}_t^n),
\end{equation}
where ${\mathbf a}_t^n=\{a_t^{k,n}| k\in{\mathcal K}\}$ and ${\mathbf b}_t^n=\{b_t^{n,m}| m\in{\mathcal C}_t^n\}$ describe the task delegation and client selection associated with FL server $s_n$ at slot $t$.
${\mathbf r}_t^n=\{r_t^{n,m}\geq 0 | m\in{\mathcal C}_t^n\}$ denote the clients' reward set within cluster ${\mathbf R}_n$, and $\mathbf{loc}_t^n=\{\text{loc}_t^m| m\in{\mathcal C}_t^n\}$ collects the locations of clients covered by FL server $s_n$ at slot $t$.
Moreover,
\begin{align}\label{eq:9}
c^n({\mathbf a}_t^n,{\mathbf b}_t^n, {\mathbf r}_t^n, \mathbf{loc}_t^n)=&\mu_1\underbrace{\left[\frac{1}{\sqrt{\frac{\tau}{\triangle_t}\left(\sum\limits_{m\in{\mathcal C}_t^n}b_t^{n,m}d_t^m \right)
}}+\frac{\triangle_t}{\tau}\right]}_{[\text{Eq.}\eqref{eq:9}-1]} \notag\\
&+\mu_2\underbrace{\left[\sum\nolimits_{k\in{\mathcal K}}a_t^{k,n}h^n+\sum\nolimits_{m\in{\mathcal C}_t^n} r_t^{n,m} \right]}_{[\text{Eq.}\eqref{eq:9}-2]},
\end{align}
accounts for the WMSFLN system cost within cluster ${\mathbf R}_n$.
$h^n$ denotes server fee for FL server $s_n$ to be paid by TR.
Inspired by \cite{zhang2022enabling}, the first term given by [Eq. \eqref{eq:9}-1] describes the expected accuracy loss for FL server $s_n$, with parameter $\mu_1$ indicating the server's valuation on accuracy loss.
Meanwhile, the second term, represented by [Eq.\eqref{eq:9}-2], denotes the TR's payment to both the FL server $s_n$'s service and the participating clients, with $\mu_2$ reflecting the importance placed on this payment.

\subsection{Queuing Model at the FL Servers}
Here, we focus on the queuing model design of the FL servers.
Before the formal introduction of virtual queue, we first present the theoretical description of FL server reputation model.
Let $g_t^n$ denote the {\em reputation value} of FL server $s_n$ at slot $t$, derived from its historical service quality.
This definition can be analogously viewed in the context of the client reputation model, as discussed in \cite{shi2023fairness}.
Different from existing works focusing on client reputation model design, we first measure the service quality of FL server to the assigned task by accuracy loss.
For notational compactness, we set $AL_t^n=[\text{Eq.}\eqref{eq:9}-1]$.
To effectively measure the quality of service of FL servers, we introduce an auxiliary variable $\sigma_t^n$, as:
\begin{align}\label{eq:11}
\sigma_t^n=\exp\left[-\frac{AL_t^{n}}{\sum\nolimits_{j\in{\mathcal N}}AL_t^j} \right].
\end{align}
If $\sigma_t^n\geq \sigma_0,$ (where $0<\sigma_0\leq 1$ is a constant), the FL server $s_n$ is considered to have delivered  positive service. Otherwise, its service is regarded as negative.
Based on the above, we deduce the FL server reputation model based on the typical Beta Reputation System (BRS) \cite{josang2002beta} and $\sigma_t^n$.
Specifically, for each FL server $s_n$, the TR tracks its positive services, denoted as $\phi_t^n$, and negative services, represented as $\psi_t^n$, over successive time slots. Thus, the dynamics of $\phi_t^n$ and $\psi_t^n$ are given by:
\begin{align}
\begin{cases}
\phi_{t+1}^n=\phi_t^n+1, \text{if~} \sigma_t^n\geq \sigma_0,\\
\psi_{t+1}^n=\psi_t^n+1, \text{otherwise}.
\end{cases}
\end{align}
At any slot $t$, BRS computes the reputation $g_t^n$ at FL server $s_n$ as
\begin{align}\label{repu}
{g_t^n}={\mathbb E}[\text{Beta}({\phi_t^n}+1, {\psi_t^n}+1)]=\frac{{\phi_t^n}+1}{{\phi_t^n}+{\psi_t^n}+2}.
\end{align}

Based on the FL server reputation model Eq. \eqref{repu}, for each FL server $s_n$, we design a virtual queue, which reflects the cumulative unfairness experienced by $s_n$ over multiple time slots. This is determined by monitoring the task delegation opportunities for $s_n$ in the context of its historical service quality.
The queuing dynamics of an FL server $s_n$'s virtual queue can be expressed as:
\begin{align}\label{eq:10}
Q_{t+1}^n=\max\left[Q_t^n+\epsilon g_t^n{\boldsymbol 1}_{[\sum_{k\in{\mathcal K}}a_t^{k,n}=0]}-\sum\nolimits_{k\in{\mathcal K}}a_t^{k,n}, ~0 \right],
\end{align}
where the term $\epsilon g_t^n{\boldsymbol 1}_{[\sum_{k\in{\mathcal K}}a_t^{k,n}=0]}\in [0, 1]$ represents the rate of unfairness accumulation for FL server $s_n$ at time slot $t$.
More precisely, $\epsilon\in[0, 1]$ is a discount factor, regulating the rate at which unfairness is generally assessed.
Without loss of generality, we set $\epsilon=K/N$ where $K$ is the number of tasks periodically generated by the TR.
To summarize, the virtual queue $Q_t^n$ increases by $\epsilon g_t^n$ when $\sum_{k\in{\mathcal K}}a_t^{k,n}=0$ during its value updates. This ensures a continuous rise in $Q_t^n$ as long $s_n$ is not chosen by the TR. On the other hand, $Q_t^n$ decreases by the value of $\sum_{k\in{\mathcal K}}a_t^{k,n}$ at any give slot $t$.
That is to say, for a FL server $s_n$ with smaller accuracy loss (i.e., higher $g_t^n$), its $Q_t^n$ grows at a higher rate if it is not delegated any FL training task.

The WMSFLN system objective is to periodically delegate FL tasks to FL servers, guided by ${\mathbf a}_t$.
Once delegated, these FL servers identify and recruit suitable clients within their wireless network coverage using ${\mathbf b}_t$ and set client rewards through ${\mathbf r}_t$.
These steps aim to reduce the overall WMSFLN cost.
However, it is essential to recognize that each client, being rational, will only agree to rewards that optimize their individual payoffs.
Thus, WMSFLN needs to offer suitable incentives to motivate clients to participate in FL training tasks. By doing so, it can also reduce the buildup of virtual queue backlogs.
The ensuing section introduces a novel framework that combines Contract Theory with Lyapunov optimization to address the conflicting objectives between the WMSFLN (i.e., TR) and clients, thereby ensuring fair task allocation despite the incomplete knowledge regarding the specific types of clients' participation.

\subsection{Contract Design}
As previously noted, the clients' participation types remain private and cannot be determined by a server with precision.
A natural framework for designing incentives that encourage clients to participate in FL training tasks is based on Contract Theory \cite{zhang2017survey}.
In our context, each FL server $s_n$ offers a contract to prospective clients within the subset ${\mathcal C}_t^n$.
The server determines this contract, establishing a correlation between the client's performance (as reflected by its type of participation), with the associated reward.
Consequently, the contract consists of a client's decision to participate and designate a specific reward, $({b_t^{n,m}, r_t^{n,m}})$, for every type $\gamma_t^{n,m}$ within ${\pmb\Gamma}$ (where ${\pmb\Gamma}$ enumerates all possible types of client participation).
Then, the TR and the FL servers determine ${\mathbf a}_t$ and ${\mathbf b}_t$ to minimize the time-average WMSFLN cost, while ensuring fair treatment of FL servers.
Once the contract is delineated, each FL server $s_n$ forwards the contract items to the clients in its coverage ${\mathcal C}_t^n$.
These clients then select the items that optimize their payoffs defined in Eq. \eqref{eq:7}.

For simplicity of expression and without loss of generality, we consider the $\Gamma$ types in an ascending sequence:  $0<\gamma_1\leq \gamma_2\leq \ldots\gamma_{\Gamma}$, encompassed within the set ${\pmb\Gamma}$.
As such, the participation of client $c_m$ under FL server $s_n$ is referred as $\gamma_i,$ if condition $\gamma_{i-1}< \gamma_t^{n,m}\leq \gamma_{i}$ holds, with $\gamma_0=0.$
In line with the Revelation Principle of Contract Theory \cite{kang2019incentive,kang2019incentivec}, for every feasible contract that accounts for private information, there exists a mechanism with equivalent payoffs that incentivizes clients to disclose their true types.
Accordingly, it is sufficient to design a contract $\pmb\Psi_t=\times_{n\in{\mathcal N}} {\pmb\Psi}_t^n$
with $\Gamma$ items, $\left\{ (b_t^{n,m}(\gamma_i), r_t^{n,m}(\gamma_i)) \right\}_{\gamma_i\in{\pmb\Gamma}}$, one for each type $\gamma_i\in{\pmb\Gamma}$, given by
\begin{align}\label{eq:13}
{\pmb\Psi}_t&=\Big\{~\left\{(b_t^{n,m}(\gamma_i), r_t^{n,m}(\gamma_i))\right\}_{\gamma_i\in{\pmb\Gamma}}~\notag\\
&| b_t^{n,m}(\gamma_i)\in{\mathbf B}_{\mathbf{loc}_t}, r_t^{n,m}(\gamma_i)\geq 0, n\in{\mathcal N}, m\in{\mathcal M} \Big\},
\end{align}
where ${\pmb\Psi}_t^n$ is a sub-contract specified by FL server $S_n$, defined as
\begin{align}\label{eq:14}
{\pmb\Psi}_t^n=&\Big\{ \left\{ \left(b_t^{n,m}(\gamma_i), r_t^{n,m}(\gamma_i) \right) \right\}_{\gamma_i\in{\pmb\Gamma}} \notag\\
&| b_t^{n,m}(\gamma_i)\in{\mathbf B}_{\mathbf{loc}_t}, r_t^{n,m}(\gamma_i)\geq 0, m\in {\mathcal C}_t^n \Big\}.
\end{align}

Upon introducing $\pmb\Gamma$, the constraint on the set ${\mathbf B}_{\mathbf{loc}_t}$ in Eq. \eqref{constraint2} can be reformulated as:
\begin{align}\label{eq:15}
\sum\nolimits_{i=1}^{\Gamma}b_t^{n,m}(\gamma_i)\leq {\boldsymbol 1}_{[a_t^n\neq 0, m\in{\mathcal C}_t^n]}, \forall n\in{\mathcal N}.
\end{align}
Then, each client $c_m\in{\mathcal C}_t^n, \forall n\in{\mathcal N}$ selects the contract item that maximizes its payoff from joining a FL task offered by server $s_n$, i.e.,
\begin{align}\label{eq:16}
&\max_{\left\{ \left(b_t^{n,m}(\gamma_i), r_t^{n,m}(\gamma_i)  \right)\right\}_{\gamma_i\in{\pmb\Gamma}}} u_t^{n,m}
\left(b_t^{n,m}(\gamma_i),r_t^{n,m}(\gamma_i) | \gamma_i \right) \notag\\
&\triangleq\max_{\left\{ \left(b_t^{n,m}(\gamma_i), r_t^{n,m}(\gamma_i)  \right)\right\}_{\gamma_i\in{\pmb\Gamma}}} (r_t^{n,m}(\gamma_i)-b_t^{n,m}(\gamma_i)/\gamma_i),
\end{align}
where the client's payoff in Eq. \eqref{eq:7} is a function of the contract item $(b_t^{n,m}(\gamma_i), r_t^{n,m}(\gamma_i))$ for the type $\gamma_i$.

While the exact types of clients' participation are not directly known to the FL servers, we follow a reasonable assumption in \cite{du2017contract,asheralieva2016optimal} that FL servers can infer the distribution of these types using historical data.
Let $\pi_i^{n,m}$ denote the probability that the participation type of client $c_m$ is $\gamma_i$ (i.e., $\pi_i^{n,m}=\Pr\{ \gamma_{i-1}<\gamma_t^{n,m}\leq\gamma_i\}
=\Pr\{  \gamma_t^{n,m}\leq \gamma_i \}-\Pr\{\gamma_t^{n,m}\leq \gamma_{i-1} \}$).
Let set ${\pmb\Pi}$ collect the distribution of all clients' participation types:
\begin{align}\label{eq:17}
\pmb\Pi=\left\{\pi_i^{n,m} | \sum\nolimits_{j=1}^{\Gamma}\pi_j^{n,m}=1, \forall m\in{\mathcal M}, n\in{\mathcal N}, 1\leq i\leq \Gamma \right\}.
\end{align}

\subsection{Contract Properties}
The two fundamental properties of Contract Theory self-revealing mechanism are individual rationality (IR) and incentive compatibility (IC). Without adhering to these principles, the honesty of clients cannot be guaranteed. The detailed definitions of IR and IC are provided subsequently.
\begin{definition}
(IR): A client only participates a training task when its payoff is non-negative, i.e.,
\begin{align}\label{eq:19}
u_t^{n,m}\left(b_t^{n,m}(\gamma_i),r_t^{n,m}(\gamma_i)| \gamma_i\right) \geq 0, \forall \gamma_i\in{\pmb\Gamma},
\end{align}
for all $n\in{\mathcal N}$ and $m\in{\mathcal M}$.
\end{definition}
\begin{definition}
(IC): A client can maximize its payoff if and only if it honestly chooses the contract item for its type, i.e.,
\begin{align}\label{eq:20}
u_t^{n,m}\left(b_t^{n,m}(\gamma_i),r_t^{n,m}(\gamma_i)| \gamma_i\right) \geq u_t^{n,m}\left(b_t^{n,m}(\gamma_j),r_t^{n,m}(\gamma_j)| \gamma_i\right),
\end{align}
for all $\gamma_i, \gamma_j\in{\pmb\Psi}_t$ and $n\in{\mathcal N}, m\in{\mathcal M}$.
\end{definition}

For a WMSFLN system, the purpose of designing this contract is to minimize its cost, while ensuring queue stability.
\begin{definition}
(Mean Rate Stable): To prevent the virtual queues from growing towards infinity, they must maintain mean rate stability throughout the FL process, i.e.,
\begin{align}\label{eq:21}
\lim_{T\to \infty}\frac{1}{T}\sum\nolimits_{t=0}^T{\mathbb E}\{Q_t^n\}=0, \forall n\in{\mathcal N}.
\end{align}
\end{definition}

Furthermore, a WMSFLN system aims to minimize expected cost and ensure virtual queue stability, with respect to the known type distribution $\pmb\Pi$.
To this end, a stochastic optimization problem for the WMSFLN, referred to as joint task delegation and client incentive mechanism (JTA-CIM) can be formulated as:
\begin{align}\label{eq:22}
\min_{({\mathbf a}, {\pmb\Psi})}&\lim_{T\to \infty} \frac{1}{T}\sum\nolimits_{t=0}^{T} {\mathbb E}\{c({\mathbf a}_t, {\pmb\Psi}_t, \mathbf{loc}_t)|{\pmb\Pi}\} \notag\\
& \eqref{eq:19}-\eqref{eq:21}, \text{~and~} {\mathbf a}_t\in {\mathbf A}_{\mathbf{loc}_t},
\end{align}
and the expected WMSFLN system cost at slot $t$ is given by:
\begin{align}\label{eq:23}
&{\mathbb E}\{c({\mathbf a}_t, {\pmb\Psi}_t, \mathbf{loc}_t)|{\pmb\Pi}\}
=\mu_2\sum\nolimits_{n\in{\mathcal N}}h^n\sum\nolimits_{k\in{\mathcal K}}a_t^{k,n}\notag \\
&+\mu_2\sum\nolimits_{n\in{\mathcal N}}\sum\nolimits_{\gamma_i\in{\pmb\Gamma}}\sum\nolimits_{m\in{\mathcal C}_t^n}\pi_i^{n,m}r_t^{n,m}(\gamma_i) \notag\\
&+\mu_1\sum_{n\in{\mathcal N}}\left[\frac{1}{\sqrt{\frac{\tau}{\triangle_t}\left(\sum\limits_{\gamma_i\in{\pmb\Gamma}}\sum\limits_{m\in{\mathcal C}_t^n}\pi_i^{n,m}b_t^{n,m}(\gamma_i)d_t^m \right)
}}+\frac{\triangle_t}{\tau}\right].
\end{align}

\subsection{Contract Feasibility and Optimality}
Contrasting with the traditional frameworks in Contract Theory, where the vale of contract items is solely dependent on the respective clients' types \cite{kang2019incentive,lim2020dynamic}, our approach, termed \methodname{}, establishes the objectives and values of contract items through a multifaceted analysis, integrating a diverse array of factors.
This novel approach considers not just the clients' inherent types, but also incorporates their current states as integral factors in the valuation process.
This methodology aligns closely with the nuanced dynamics observed in FL environments, where static characteristics (client types), dynamic factors (current states $\mathbf {loc}_t$, task delegation ${\mathbf a}_t$), and the virtual queue stability constraints outlined in Eq. \eqref{eq:21}  critically influence the efficacy and relevance of contractual agreements.

Due to the coupling diverse factors, the resulting contract departs from the classical framework. In this setting, we need to redefine the conditions for the feasibility and optimality of \methodname{} contracts. Based on [Property 1 of \cite{lu2021toward}], we derive the necessary and sufficient conditions to guarantee contract feasibility, presented in Lemma \ref{lemma1} below.
\begin{lemma}\label{lemma1}
A feasible contract $\pmb\Psi_t$ must meet the following necessary and sufficient conditions:
\begin{subequations}\label{add:1}
\begin{numcases}{}
b_t^{n,m}(\gamma_1)\leq b_t^{n,m}(\gamma_2)\leq \cdots\leq b_t^{n,m}(\gamma_{\Gamma}), \notag\\
r_t^{n,m}(\gamma_1)\leq r_t^{n,m}(\gamma_2)\leq\cdots\leq r_t^{n,m}(\gamma_{\Gamma}),   \label{add:1a}\\
r_t^{n,m}(\gamma_1)-b_t^{n,m}(\gamma_1)/\gamma_1\geq 0, \label{add:1b}\\
b_t^{n,m}(\gamma_{i-1})+\gamma_{i-1}(r_t^{n,m}(\gamma_i)-r_t^{n,m}(\gamma_{i-1})) \notag\\
\leq b_t{n,m}(\gamma_{i})\notag\\
\leq b_t^{n,m}(\gamma_{i-1})+\gamma_{i}(r_t^{n,m}(\gamma_i)-r_t^{n,m}(\gamma_{i-1})), \label{add:1c}
\end{numcases}
\end{subequations}
for all $\gamma_i\in{\pmb\Psi_t}, m\in{\mathcal M}, n\in{\mathcal N}$.
\end{lemma}

Having established the preliminary conditions necessary for contract feasibility in our framework, we are now present the detailed design of an optimal contract, as outlined in Theorem \ref{theorem1}. We refer the interested readers to \cite{zhang2022enabling} for a detailed proof of Theorem \ref{theorem1}.
\begin{theorem}\label{theorem1}
A unique optimal contract $\overline{\pmb\Psi}_t$ for problem \eqref{eq:22} is described by the following equations:
\begin{align}\label{eq:25}
\begin{cases}
\bar{b}_t^{n,m}(\gamma_1)&=\cdots=\bar{b}_t^{n,m}(\gamma_{\Gamma-1})=0, \bar{b}_t^{n,m}(\gamma_{\Gamma})=1, \\
\bar{r}_t^{n,m}(\gamma_1)&=\cdots=\bar{r}_t^{n,m}(\gamma_{\Gamma-1})=0,\bar{r}_t^{n,m}(\gamma_{\Gamma})=\frac{1}{\gamma_{\Gamma}},
\end{cases}
\end{align}
for all $n\in{\mathcal N}, m\in{\mathcal M}$.
\end{theorem}

According to Theorem \ref{theorem1}, in any optimal contract, only the contract items related to the highest participation type, $\gamma_{\Gamma}$, have positive values. All other items are set to zero.
Substituting the solution of Theorem \ref{theorem1} into Eq. \eqref{eq:22} yields
\begin{align}\label{eq:26}
\min_{({\mathbf a}, {\pmb\Psi}^b)}&\lim_{T\to\infty}
\frac{1}{T}\sum\nolimits_{t=1}^T{\mathbb E}\{c({\mathbf a}_t, {\pmb\Psi}_t^b, \mathbf{loc}_t)|{\pmb\Pi}\} \notag\\
&\text{s.t.~} \eqref{eq:21} \text{~and~} {\mathbf a}_t\in{\mathbf A}_{\mathbf{loc}_t}.
\end{align}
In the problem \eqref{eq:26}, the optimal contracts defined by ${\pmb\Psi}^b=\{{\pmb\Psi}_t^b\}_{t\in{\mathbb N}_0}=\left\{b_t^{n,m}(\gamma_{\Gamma})\in \widetilde{\mathbf B}_{\mathbf{loc}_t} | n\in{\mathcal N}, m\in{\mathcal M}\right\}_{t\in{\mathbb N}_0}$, where
\begin{align}\label{eq:27}
\widetilde{\mathbf B}_{\mathbf{loc}_t}=\{b_t^{n,m}(\gamma_{\Gamma})\in\{0, 1\} \mid &b_t^{n,m}(\gamma_{\Gamma})\leq {\boldsymbol 1}_{[\sum_{k\in{\mathcal K}}a_t^{k,n}\neq 0]}, \notag \\
& n\in{\mathcal N}, m\in{\mathcal M} \}.
\end{align}
The expected WMSFLN system cost ${\mathbb E}\{c({\mathbf a}_t, {\pmb\Psi}_t^b, \mathbf {loc}_t)|{\pmb\Pi}\}$ at slot $t$ can be derived by considering only the highest participation type $\gamma_{\Gamma}$ in Eq. \eqref{eq:23}.


\subsection{Lyapunov Optimization}
It is noted that solving the equivalent optimization problem in Eq. \eqref{eq:26} is challenging due to the lack of prior information on clients' distributions, queue backlogs, and clients' participation costs.
The widely used Lyapunov optimization techniques \cite{neely2022stochastic} are commonly referred to as effective methods to solve time averaged constraints.
Likewise, we first define a control policy $({\mathbf a}, {\pmb\Psi}^b)$ deduced from optimization Eq. \eqref{eq:26}.
In this framework, our aim is to formulate a control policy $(\bar{\mathbf a}, \bar{\pmb\Psi}^b)$-essentially a sequence of control policies $\{(\bar{\mathbf a}_t, \bar{\pmb\Psi}_t^b)\}_{t\in{\mathbb N}_0}$, which reduces the time-average WMSFLN cost and ensures queue stability without prior knowledge of the mentioned statistics.
We then employ Lyapunov optimization techniques to ensure that every increment of $Q_t^n$ adheres to Eq. \eqref{eq:21}  as effectively as possible.
Let ${\pmb\Theta}_t=\{Q_t^1, Q_t^2, \ldots, Q_t^N\}$.
Consider a quadratic Lyapunov function \cite{neely2013dynamic}, $L(\Theta_t)$:
\begin{align}\label{eq:29}
L(\pmb\Theta_t)=\frac{1}{2}\sum\nolimits_{n=1}^N (Q_t^n)^2 \geq 0.
\end{align}
It serves as an indicator of unfairness delegation among FL servers.
A lower value of $L(\pmb\Theta_t)$ suggests that there is minimal and evenly spread unfairness across the servers.

To measure the expected increase of $L({\pmb\Theta}_t)$ in one time step, we formulate the Lyapunov drift, $\triangle L({\pmb\Theta}_t)$, expressed as $\triangle L({\pmb\Theta}_t)={\mathbb E}\left\{L({\pmb\Theta}_{t+1})-L({\pmb\Theta}_t) |{\pmb\Theta}_t \right\}.$
By minimizing $\triangle L({\pmb\Theta}_t)$, we aim to limit the growth of the all $Q_t^n$ by dynamically task delegation among FL servers.
$\triangle L({\pmb\Theta}_t)$ can be further derived into
\begin{align}\label{eq:31}
\triangle L({\pmb\Theta}_t)\leq \sum_{n=1}^N \left[Q_t^n\left(\epsilon g_t^n{\boldsymbol 1}_{[\sum_{k\in{\mathcal K}} a_t^{k,n}=0]}-\sum_{k\in{\mathcal K}}a_t^{k,n}\right)+\theta\right],
\end{align}
where $\theta=1.$ The proof process is detailed in the following.

\begin{proof}
Based on \eqref{eq:13} and \eqref{eq:29}, we have
\begin{align}\label{eq:36}
\triangle L({\pmb\Theta}_t)=&L({\pmb\Theta}_{t+1})-L({\pmb\Theta}_t) \notag\\
=&\frac{1}{2}\sum\nolimits_{n\in{\mathcal N}}\left[(Q_{t+1}^n)^2-(Q_t^n)^2\right] \notag\\
=&\frac{1}{2}\sum_{n\in{\mathcal N}}\left\{\left(Q_t^n+\epsilon g_t^n{\boldsymbol 1}_{[\sum_{k\in{\mathcal K}} a_t^{k,n}=0]}-\sum_{k\in{\mathcal K}}a_t^{k,n}\right)^2\right. \notag\\
&\left.-(Q_t^n)^2 \right\}\notag \\
=&\frac{1}{2}\sum_{n\in{\mathcal N}}\left(\epsilon g_t^n{\boldsymbol 1}_{[\sum\nolimits_{k\in{\mathcal K}}a_t^{k,n}=0]}-\sum\nolimits_{k\in{\mathcal K}}a_t^{k,n}\right)^2 \notag\\
&+\sum_{n\in{\mathcal N}}Q_t^n\left( \epsilon g_t^n{\boldsymbol 1}_{[\sum\nolimits_{k\in{\mathcal K}}a_t^{k,n}=0]}-\sum\nolimits_{k\in{\mathcal K}}a_t^{k,n} \right) \notag\\
\leq&N\theta+\sum_{n\in{\mathcal N}}Q_t^n\left(\epsilon g_t^n{\boldsymbol 1}_{[\sum\nolimits_{k\in{\mathcal K}}a_t^{k,n}=0]}-\sum_{k\in{\mathcal K}}a_t^{k,n} \right),
\end{align}
where $\theta=\frac{1}{2}\left(\epsilon g_t^n{\boldsymbol 1}_{[\sum_{k\in{\mathcal K}}a_t^{k,n}=0]}\right)_{\max}^2+\frac{1}{2}\left(\sum_{k\in{\mathcal K}}a_t^{k,n}\right)_{\max}^2$ with
$\left(\sum_{k\in{\mathcal K}}a_t^{k,n}\right)_{\max}=1$
and $\left(\epsilon g_t^n{\boldsymbol 1}_{[\sum_{k\in{\mathcal K}}a_t^{k,n}=0]}\right)_{\max}=1$ are the upper limits for $\sum_{k\in{\mathcal K}}a_t^{k,n}$ and $\epsilon g_t^n{\boldsymbol 1}_{[\sum_{k\in{\mathcal K}}a_t^{k,n}=0]}$  for all $n\in{\mathcal N}$.
\end{proof}

Based on the above formulation, rather than minimizing $\triangle L({\pmb\Theta}_t)$, we concentrate on minimizing the drift-plus-cost function:
\begin{align}\label{eq:32}
V {\mathbb E}\{c({\mathbf a}_t, {\pmb\Psi}_t^b, \mathbf {loc}_t)|{\pmb\Pi}, {\pmb\Theta}_t\}+\triangle L({\pmb\Theta}_t),
\end{align}
where the balance parameter $V\geq 0$ is selected to strike the desired balance between WMSFLN cost and queue stability.
Based on Eq. \eqref{eq:31}, at any time slot $t$, Eq. \eqref{eq:32} is upper bounded by:
\begin{align}\label{eq:33}
V &{\mathbb E}\{c({\mathbf a}_t, {\pmb\Psi}_t^b, \mathbf {loc}_t)|{\pmb\Pi}, {\pmb\Theta}_t\}+\triangle L({\pmb\Theta}_t)\notag\\
\leq& N\theta+V {\mathbb E}\{c({\mathbf a}_t, {\pmb\Psi}_t^b, \mathbf{loc}_t)|{\pmb\Pi}, {\pmb\Theta}_t\} \notag\\
&+\sum\nolimits_{n=1}^N Q_t^n\left[\epsilon g_t^n{\boldsymbol 1}_{[\sum_{k\in{\mathcal K}} a_t^{k,n}=0]}-\sum\nolimits_{k\in{\mathcal K}}a_t^{k,n}\right].
\end{align}

According to Eq. \eqref{eq:33}, we can minimize the upper bound in Eq. \eqref{eq:33}, rather than directly minimizing the drift-plus-cost function from Eq. \eqref{eq:32}.
At any given slot $t$, upon observing the current queue ${\pmb\Theta}_t$ and clients' locations $\mathbf {loc}_t$, we select the control action $(\bar{\mathbf a}_t, \overline{\pmb\Psi}_t^b)$ (i.e.,  the optimal action) to minimize the value on the right-hand side of Eq. \eqref{eq:33}.
Furthermore, for any given ${\pmb\Theta}_t$ and $\mathbf{loc}_t$, only two terms, ${\mathbb E}\{c({\mathbf a}_t, {\pmb\Psi}_t^b, {\mathbf l}_t)|{\pmb\Pi}, {\pmb\Theta}_t\}$ and $\sum_{n=1}^N Q_t^n[\epsilon g_t^n{\boldsymbol 1}_{[\sum_{k\in{\mathcal K}}a_t^{K,n}=0]}-\sum_{k\in{\mathcal K}}a_t^{k,n}]$, are pertinent in relation to $({\mathbf a}_t, {\pmb\Psi}_t^b)$.
Consequently, we derive the following optimization problem:
\begin{align}\label{eq:34}
\min_{({\mathbf a}_t, {\pmb\Psi}_t^b)}&{\pmb\Omega}({\mathbf a}_t, {\pmb\Psi}_t^b, \mathbf{loc}_t)|{\pmb\Pi}, {\pmb\Theta}_t)\notag\\
&\text{s.t.~} {\mathbf a}_t\in{\mathbf A}_{\mathbf {loc}_t},
\end{align}
where the objective function ${\pmb\Omega}({\mathbf a}_t, {\pmb\Psi}_t^b, \mathbf{loc}_t)|{\pmb\Pi}, {\pmb\Theta}_t)$ is detail expressed as:
\begin{align}\label{eq:35}
{\pmb\Omega}&({\mathbf a}_t, {\pmb\Psi}_t^b, \mathbf {loc}_t)|{\pmb\Pi}, {\pmb\Theta}_t)=\sum\nolimits_{n\in{\mathcal N}}{\pmb\Omega}^n({\mathbf a}_t^n, {\pmb\Psi}_t^{b(n)}, \mathbf {loc}_t^n)|{\pmb\Pi}, {\pmb\Theta}_t^n)\notag\\
=&V {\mathbb E}\{c({\mathbf a}_t, {\pmb\Psi}_t^b, \mathbf{loc}_t)|{\pmb\Pi}, {\pmb\Theta}_t\} \notag\\
&+\sum_{n=1}^N Q_t^n\left[\epsilon g_t^n{\boldsymbol 1}_{[\sum_{k\in{\mathcal K}}a_t^{k,n}=0]}-\sum\nolimits_{k\in{\mathcal K}}a_t^{k,n}\right] \notag
\end{align}
\begin{align}
=&\sum\nolimits_{n\in{\mathcal N}} \left\{(\mu_2Vr^n-Q_t^n)\sum_{k\in{\mathcal K}}a_t^{k,n}+Q_t^n\epsilon g_t^n{\boldsymbol 1}_{[\sum_{k\in{\mathcal K}}a_t^{k,n}=0]}  \right.\notag\\
&\left.+V\mu_1\bigg(\frac{1}{\sqrt{{\tau}/{\triangle_t}\left(\sum_{m\in{\mathcal C}_t^n}\pi_{\Gamma}^{n,m}b_t^{n,m}(\gamma_{\Gamma})d_t^m \right)
}}+{\triangle_t}/{\tau}\bigg) \right. \notag\\
&\left.+V\mu_2\sum\nolimits_{m\in{\mathcal C}_t^n}\pi_{\Gamma}^{n,m}b_t^{n,m}(\gamma_{\Gamma})/\gamma_{\Gamma}\right\}.
\end{align}

Note that Eq. \eqref{eq:34} can be decomposed into $N$ small sub-problems that can run in parallel, as defined by:
\begin{align}\label{pro}
\min_{({\mathbf a}_t^n, {\pmb\Psi}_t^{b(n)})}&{\pmb\Omega}^n({\mathbf a}_t^n, {\pmb\Psi}_t^{b(n)}, \mathbf {loc}_t^n)|{\pmb\Pi}, {\pmb\Theta}_t^n)\notag\\
&\text{s.t.~} {\mathbf a}_t^n\in {\mathbf A}_{\mathbf{loc}_t^n},
\end{align}
for all $n\in{\mathcal N}$.
In Eq. \eqref{pro}, ${\pmb\Theta}_t^n$ represents the history  queue backlogs within the region ${\mathbf R}_n$, ${\pmb\Psi}_t^{b(n)}=\{b_t^{n,m}(\gamma_{\Gamma})\in\{0,1\}| b_t^{n,m}(\gamma_{\Gamma})\leq {\mathbf 1}_{[\sum_{k\in{\mathcal K}}a_t^{k,n}=1]} | m\in{\mathcal C}_t^n\}$ is the contract assigned by FL server $s_n$ at slot $t$.
The problem in Eq. \eqref{pro} can be addressed independently by each FL server,  as it relies solely on locally observable data, specifically the queue backlog history ${\pmb\Theta}_t^n$ and the state $\mathbf{loc}_t^n$, without the need for information exchange with other servers.
At the start of each time slot $t$, every FL server observes the historical ${\pmb\Theta}_t^n$ and state $\mathbf{loc}_t^n$. Then, it selects the control policy $(\bar{\mathbf a}_t^n, \bar{\pmb\Psi}_t^{b(n)})$ that solves the problem \eqref{pro} and outputs the results.
\begin{figure*}[!t]
\centering
\subfigure[]{
\begin{minipage}[t]{0.45\linewidth}
\includegraphics[width=3.5in]{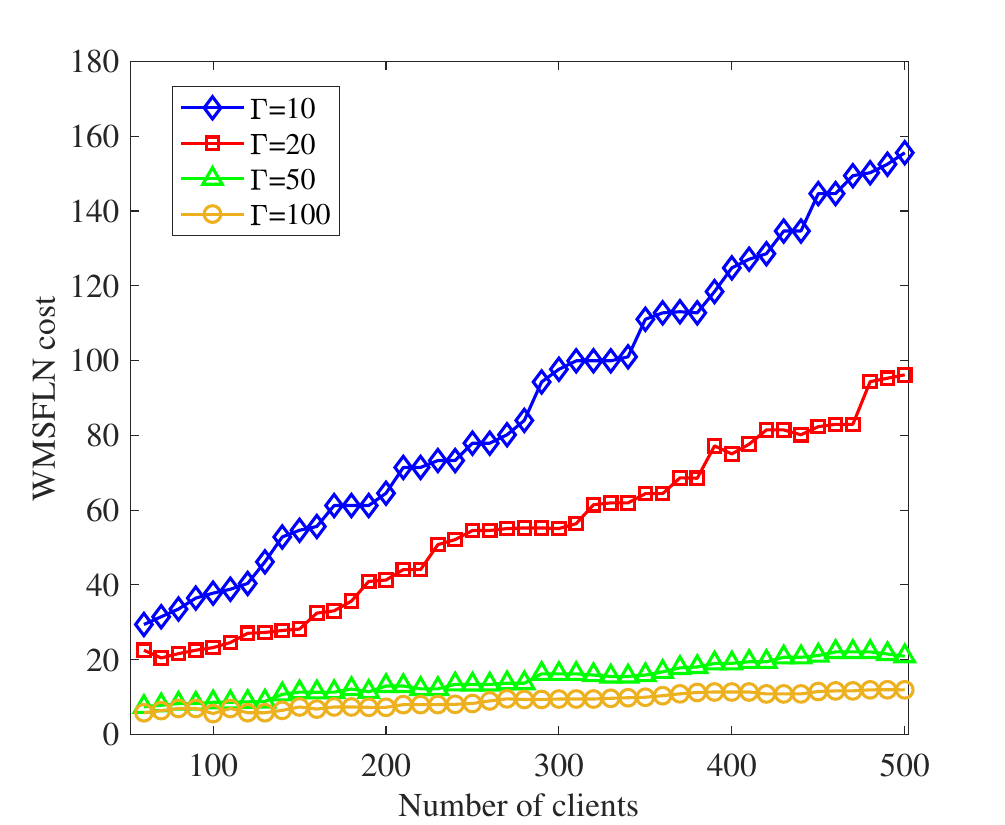}
\end{minipage}
\label{fig:2a}
}
\subfigure[]{
\begin{minipage}[t]{0.45\linewidth}
\includegraphics[width=3.5in]{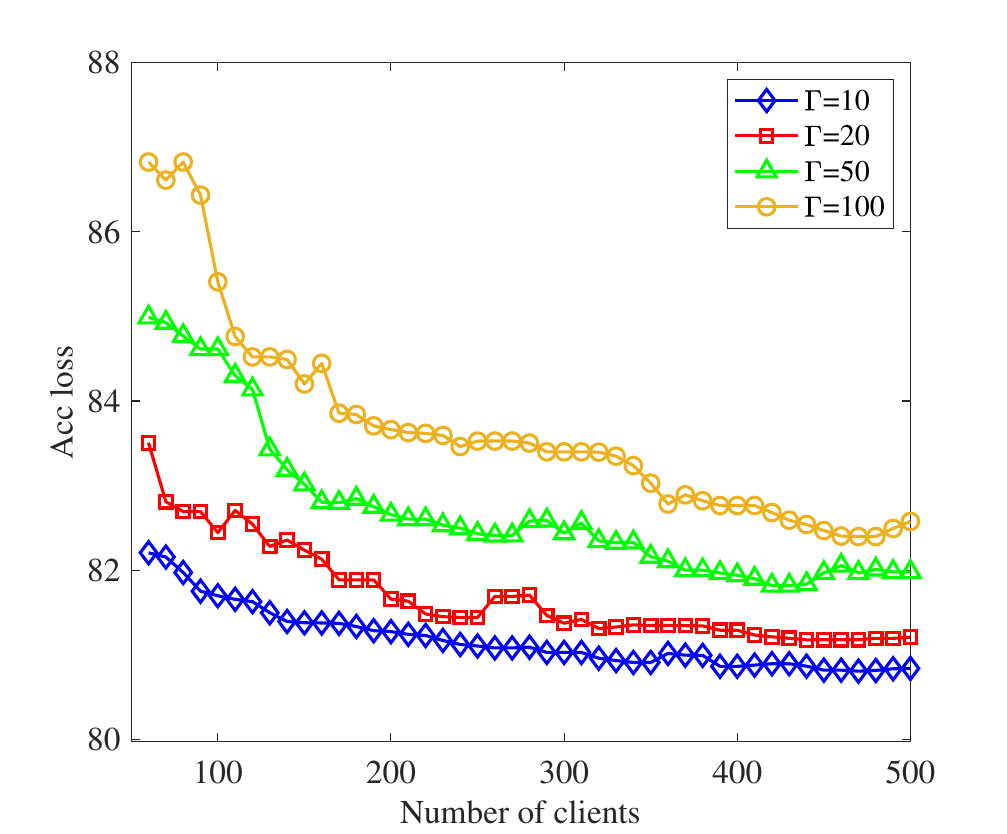}
\end{minipage}
\label{fig:2b}
}
\caption{(a) Time-averaged WMSFLN cost and (b) time-averaged accuracy loss versus the number of clients $M$ using $\Gamma=10, 20, 50, 100$, for $N=10$ and $V=10$.}
\end{figure*}

\section{Performance Evaluation}
\subsection{Evaluation Setup}
\subsubsection{WMSFLN System Settings}
To keep the complexity of the simulations tractable while considering a significantly loaded system, we assume that $10$ FL servers are deployed in a two-dimensional $(100\times 200)\text{~m}$ rectangular area following an uniform distribution.
That is to say, each FL server is located in an equal-size rectangular area, where the size is $(50\times 40)\text{~m}$.
The initial locations of $200$ clients satisfy a homogeneous Poisson Point Process (PPP) \cite{andrews2011tractable,zheng2018stackelberg}, meaning each client is independently and uniformly located in this area.
To accurately represent a realistic mobile environment, we employ the Gauss-Markov Mobility Model (GMMM) \cite{gao2019dynamic}, to simulate the movements of mobile clients.


In our analysis, the channel realizations are generated according to the 3GPP propagation environment \cite{hoglund20183gpp}.
Moreover, we adapt Rayleigh fading to simulate multipath effects. Throughout the experiments, we consistently adopt the relevant parameters as delineated in [Table I of \cite{gao2019dynamic}].
Regarding the WMSFLN parameters, we adopt $K=8, \mu_1=0.1, \mu_2=0.9, \tau=1\text{~s}, \triangle_t=0.1\text{~s}$, and $\Gamma=20$.
\begin{figure*}[!t]
\centering
\subfigure[]{
\begin{minipage}[t]{0.45\linewidth}
\includegraphics[width=3.5in]{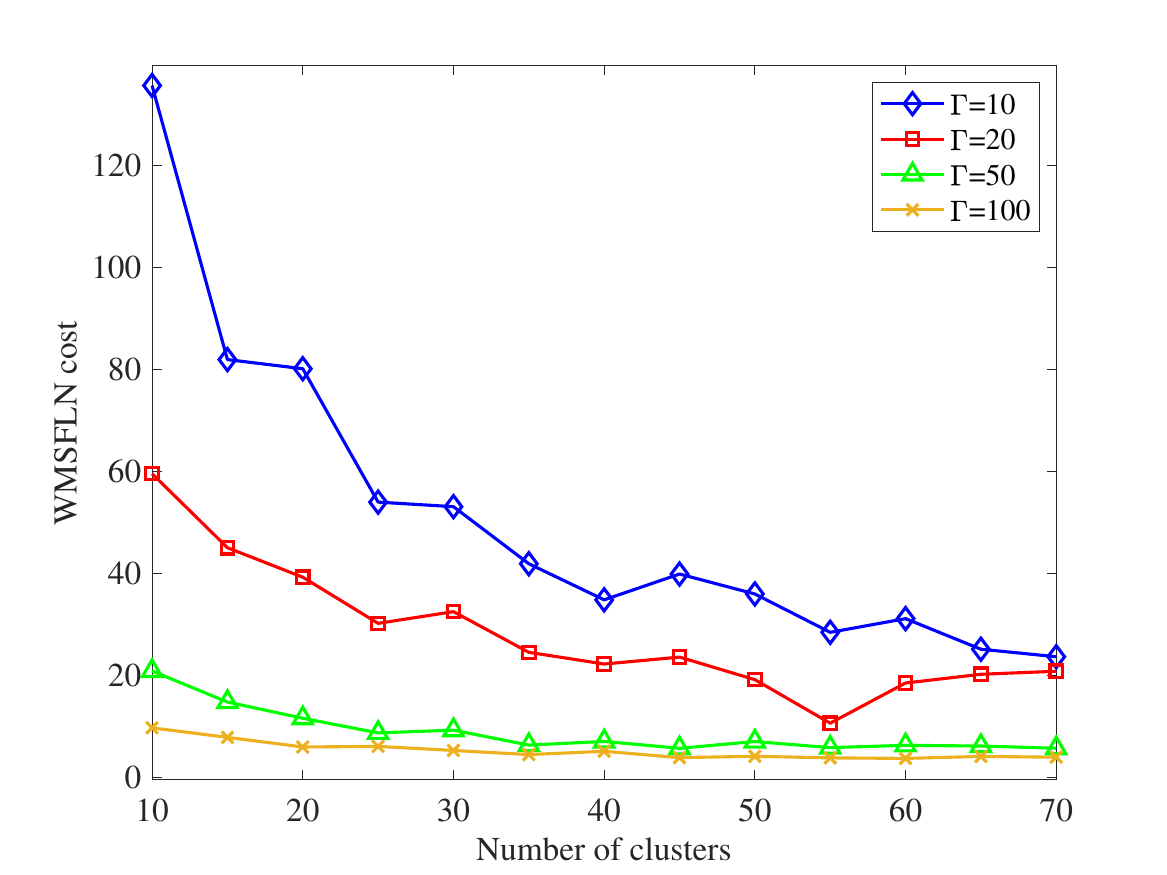}
\end{minipage}
\label{fig:3a}
}
\subfigure[]{
\begin{minipage}[t]{0.45\linewidth}
\includegraphics[width=3.5in]{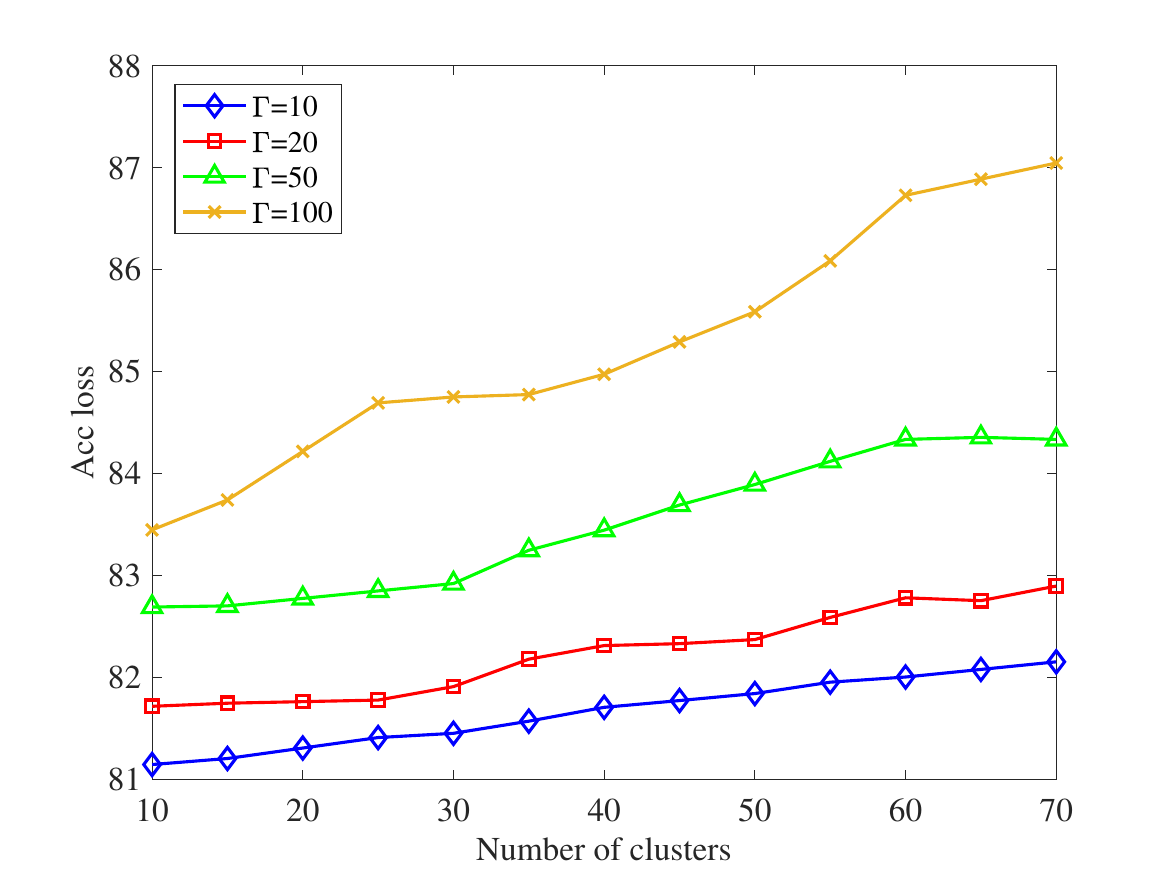}
\end{minipage}
\label{fig:3b}
}
\caption{Time-averaged WMSFLN cost and (b) time-averaged accuracy loss versus the number of FL servers $N$ using $\Gamma=10, 20, 50, 100$, for $M=500$ and $V=10$.}
\end{figure*}
\begin{figure*}[!t]
\centering
\subfigure[]{
\begin{minipage}[t]{0.45\linewidth}
\includegraphics[width=3.5in]{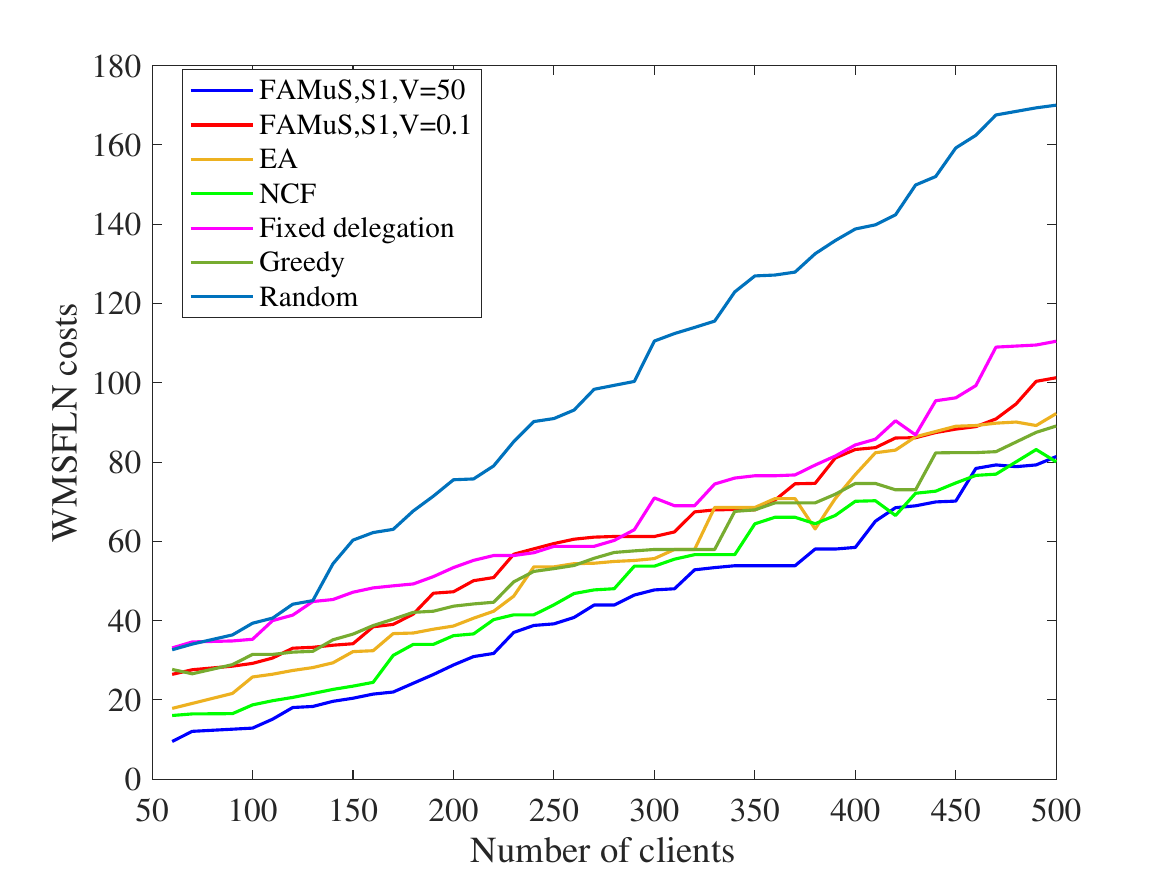}
\end{minipage}
\label{fig:4a}
}
\subfigure[]{
\begin{minipage}[t]{0.45\linewidth}
\includegraphics[width=3.5in]{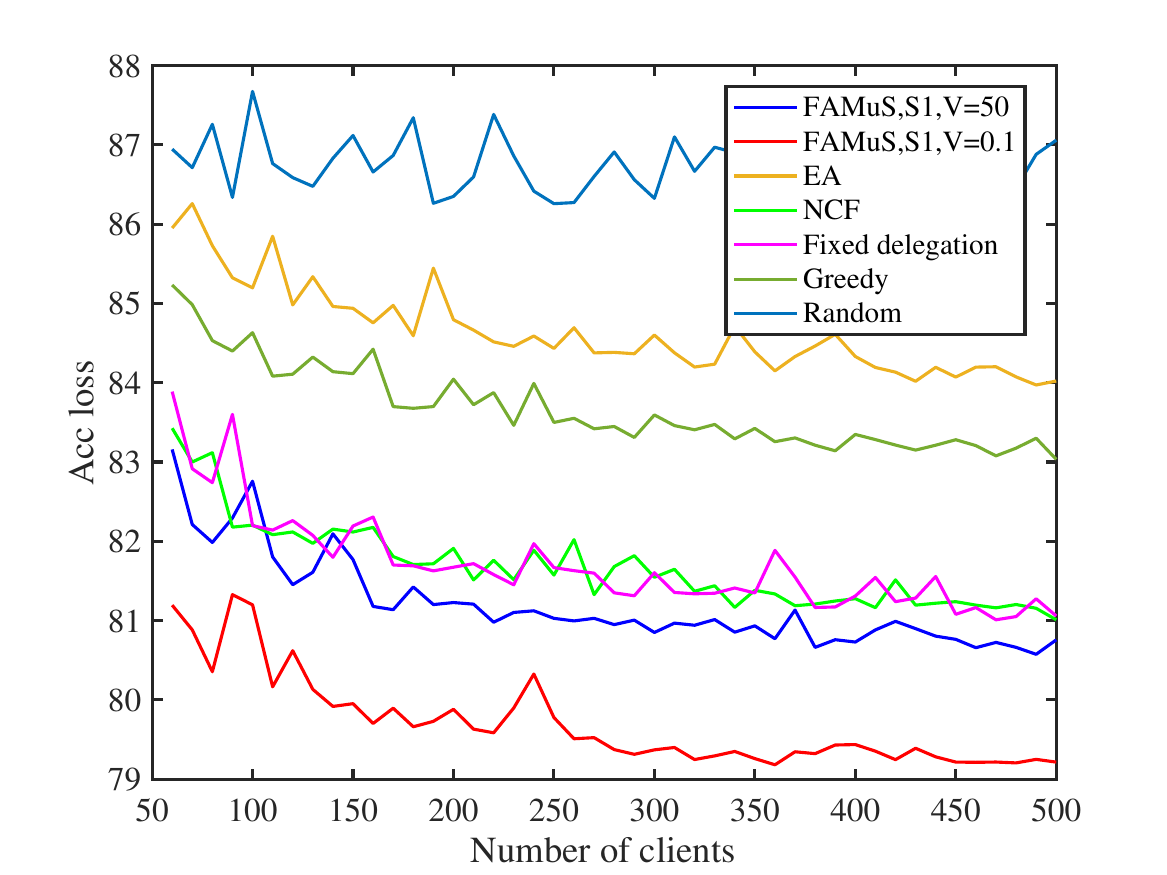}
\end{minipage}
\label{fig:4b}
}
\label{fig:4}
\caption{(a) Time-averaged WMSFLN cost performance comparison and (b) accuracy loss versus the number of clients $M$ for $N=10$ and $\Gamma=50$ in Scenario 1. }
\end{figure*}
\subsubsection{Comparison Baselines}
In order to make sufficient comparisons, we select the following five relevant baselines to compare with our proposed \methodname{} approach.
\begin{enumerate}
    \item{\bf Random}: both the delegated FL servers and the participants are randomly determined.
    \item{\bf Greedy}: the delegated FL servers and the participants are optimized for unilateral cost minimization.
    \item {\bf NCF} \cite{lu2021toward}: which only focuses on minimizing cost without taking any fairness into account. In order to deploy this method into the WMSFLN, the task delegation is derived by a greedy algorithm.
    \item {\bf EA} \cite{mcmahan2017communication}: which assigns the same number of tasks to all participants, ignoring fairness and the heterogeneity of clients. In order to apply this method in WMSFLN, the delegated probability of each FL server is $K/N$ at any slot.
    \item {\bf Fixed Delegation:} the delegated FL servers  remain constant across all scenarios.
\end{enumerate}

In addition, we consider two contract scenarios:
\begin{enumerate}
    \item {\bf Scenario 1}: the same setting as the \methodname{}, i.e., the contract is periodically updated;
    \item {\bf Scenario 2}: a uniform contract as benchmark, which contains a single uniform contract item for all clients.
\end{enumerate}

Furthermore, we utilize Jain's Fairness Index (JFI) \cite{jain1984quantitative} to assess the fairness of task delegation achieved by each approach upon reaching FL model convergence.
It is formulated as follows:
\begin{align}\label{eq:39}
\text{JFI}=\frac{\left( \sum_{n=1}^N{x^n}/{\sigma^n}\right)^2}
{N\times \sum_{n=1}^N\left({x^n}/{\sigma^n} \right)^2 },
\end{align}
where $x^n$ represents the total number of delegation time for server $s_n$ by the end of the training.
$\sigma^n$ represents the accumulated service quality by the end of the training.
\begin{figure*}[!t]
\centering
\subfigure[]{
\begin{minipage}[t]{0.45\linewidth}
\includegraphics[width=3.5in]{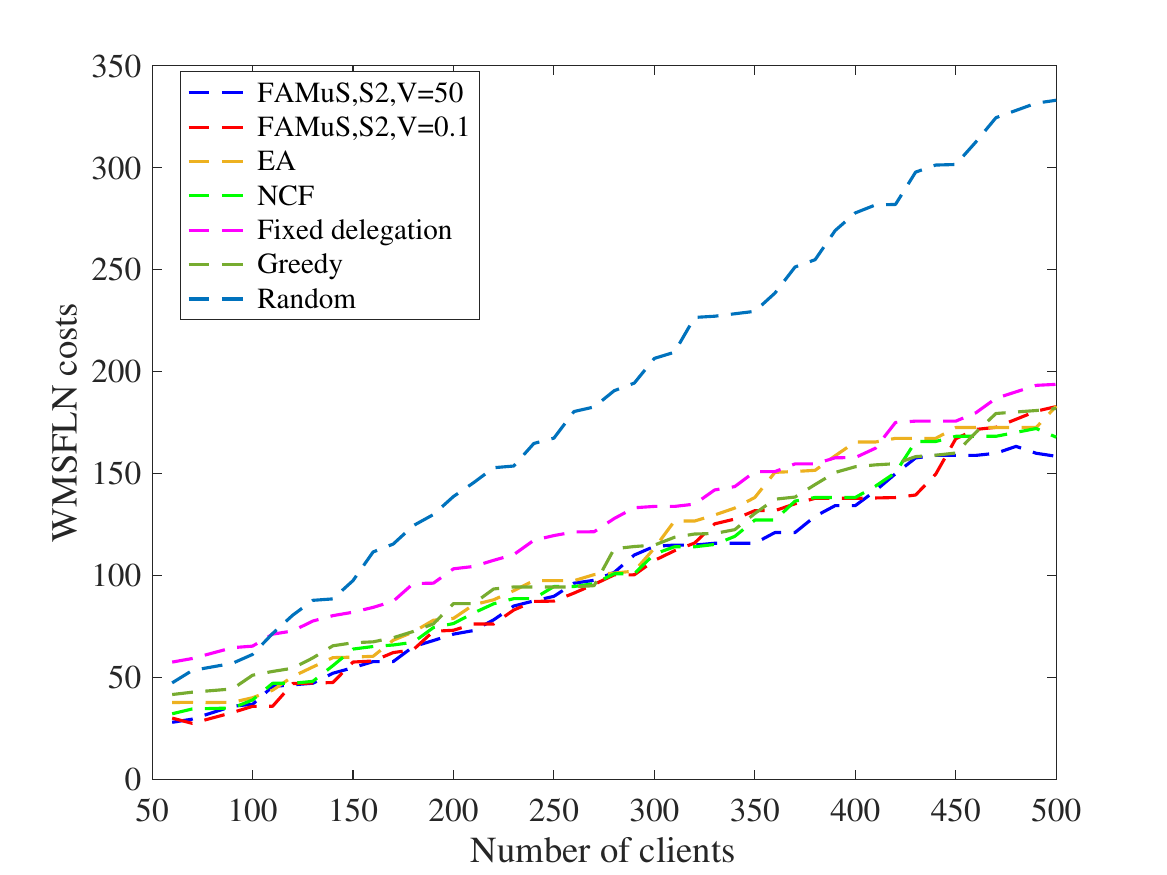}
\end{minipage}
\label{fig:5a}
}
\subfigure[]{
\begin{minipage}[t]{0.45\linewidth}
\includegraphics[width=3.5in]{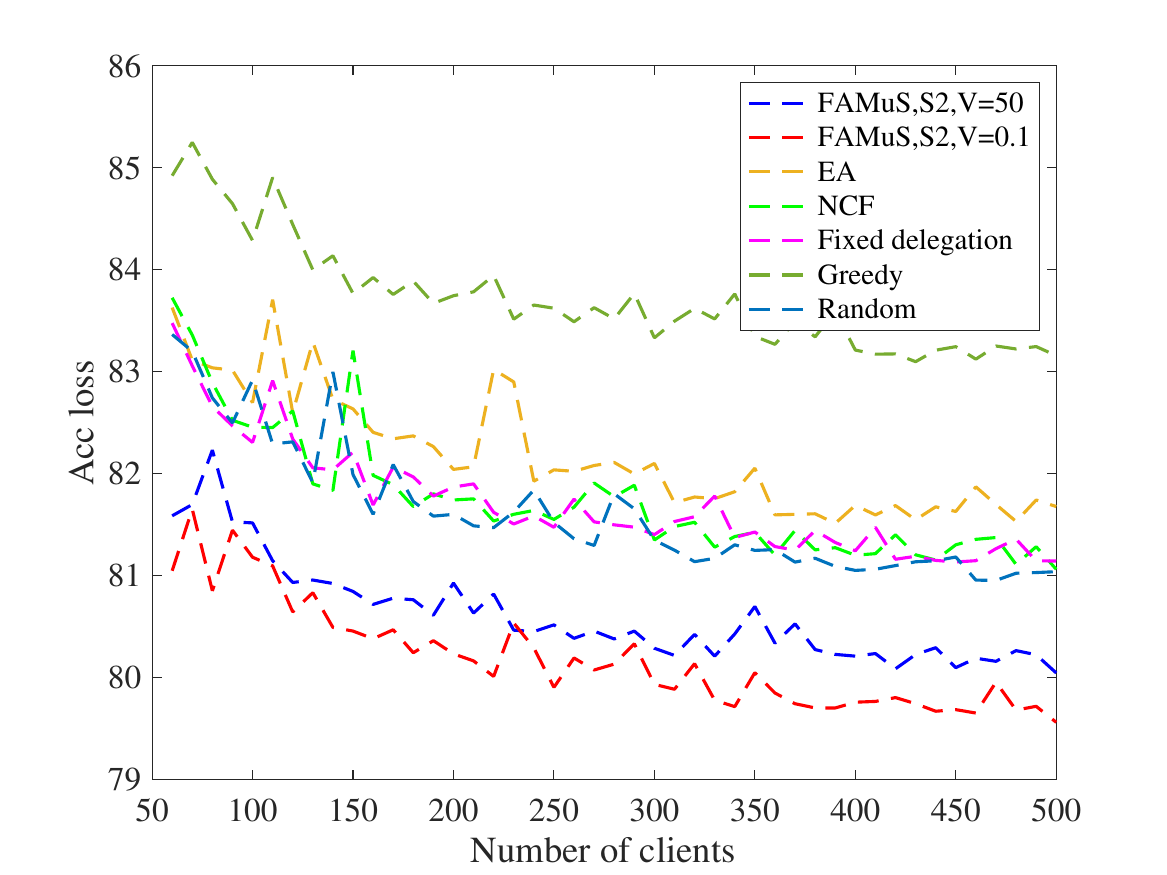}
\end{minipage}
\label{fig:5b}
}
\label{fig:5}
\caption{(a) Time-averaged WMSFLN cost performance comparison and (b) accuracy loss versus the number of clients $M$ for $N=10$ and $\Gamma=50$ in Scenario 2.}
\end{figure*}

\subsubsection{FL Implementation}
Note that our proposed \methodname{} approach for FL in the WMSFLN system can be regarded as a pre-selection component integrated with existing federated platforms.
We establish an FL environment using PyTorch and employ two different convolutional neural network models for classification tasks on MNIST and CIFAR-10 datasets.
For CIFAR-10, the training size $d_m$ for each client is randomly selected from $[400, 500]\text{~MB}$.
Similarly, for MNIST, the training size $d_m$ falls within the range of $[100, 200]\text{~MB}$ for each client.

In the FL setup, the local training parameters are set as follows: The global epoch is $50$ and the number of local epoch is $10$.
Each local epoch consists of $10$ batches, with a learning rate of $0.01$ and Stochastic Gradient Descent (SGD) momentum of $0.5$.
Two distinct Convolutional Neural Network (CNN) architectures are employed for the MNIST and CIFAR-10 datasets.
The MINIST CNN (MnistCNN) consists of two convolutional layers with dropout, two fully connected layers, and employs ReLU activation and max pooling, outputting through a log softmax function. In contrast, the CIFAR CNN (CifarCNN) features two convolutional layers followed by max pooling, three linear layers, and also uses ReLU activation, culminating in a log softmax output.
Both architectures are tailored to efficiently process their respective datasets, leveraging convolution, pooling, and activation techniques to optimize performance.

\begin{table*}[!t]
  \centering
  \caption{Performance comparison under Scenario 1 (S1) and Scenario 2 (S2), with $V=10$.}
  \label{table:1}
  \setlength{\tabcolsep}{0.9mm}{
\begin{tabular}{|c|c|c|c|c|c|c|c|c|c|c|c|c|}
\hline
\multirow{3}{*}{Method}&\multicolumn{6}{c|}{MNIST}&\multicolumn{6}{c|}{CIFAR-10} \\
\cline{2-7} \cline{8-13}
& \multicolumn{3}{c|}{S1} &\multicolumn{3}{c|}{S2} &\multicolumn{3}{c|}{S1} & \multicolumn{3}{c|}{S2}\\
\cline{2-4}  \cline{5-7} \cline{8-10} \cline{11-13}
& Cost & Acc (\%) & Fairness & Cost & Acc (\%) & Fairness & Cost & Acc (\%) &Fairness& Cost & Acc (\%) & Fairness \\
\hline
Random              & 33.41&77.64&0.871&127.38&70.39&0.861&103.87&54.79&0.869&146.12&53.49&0.873\\
Greedy              & 40.44&78.93&0.876&91.24&73.96&0.867&61.10&55.55&0.865&89.84&54.68&0.877\\
EA                  & \underline{22.46} & 80.78 & \underline{0.889} & 86.04 & 75.19 & \underline{0.893} &66.31&56.34&\underline{0.890} &\underline{82.74} &54.39& \underline{0.894}\\
NCF                 & 22.14 & 80.52 & 0.800 & \underline{82.50} & 75.09 & 0.800& \underline{57.24} & 56.19 &0.800 &{\bf 80.66} &54.09 &0.800\\
Fixed Delegation    & 79.51 & \underline{81.67} & 0.800 & 88.37 & \underline{76.86}& 0.800 &62.77 & \underline{58.75} & 0.800 & 94.11 & \underline{55.89} & 0.800 \\
\hline
\methodname{}       & {\bf 13.29} &{\bf 89.75}&  {\bf 0.899} & {\bf 79.00} & {\bf 84.86} & {\bf 0.897} & {\bf 13.55} & {\bf 65.44} & {\bf 0.900} & { 89.67} & {\bf 60.78} & {\bf 0.895}\\
\hline
   \end{tabular}}
\end{table*}
\subsection{Results and Discussion}
To demonstrate the performance of the proposed \methodname{}, we depict in Figures \ref{fig:2a} and \ref{fig:2b} the time-average WMSFLN cost and accuracy loss as function of the number of clients $M$ for different numbers of types $\Gamma$.
As can be seen, the WMSFLN cost decreases as the number of types increases but increases with an increasing number of clients.
By contrast,  the accuracy loss increases as the number of types, $\Gamma,$ increases, but decreases as $M$ increases.
This in essence attributes to the fact that a decrease in the number of types corresponds to a reduction in the value of the highest type.
Consequently, this leads to a heightened participation cost for clients, thereby contributing to an increased WMSFLN cost.
Notably, with regard to accuracy loss, a reduced number of types facilitates the recruitment of high-quality of participants.
This aligns with the prior observation that a few high-quality participants often outperform a large number of staggers in FL.

In Figures \ref{fig:3a} and \ref{fig:3b}, for a given number of clients, $M$, we evaluate the time-averaged WMSFLN cost and accuracy loss versus the number of clusters, respectively.
From the results, we observe that the time-averaged WMSFLN cost decreases, when $\Gamma$ increases.
However, for a given value of $\Gamma$, the WMSFLN cost performance is slightly improved, when increasing the value of $N$.
As seen in Figure \ref{fig:3b}, the performance of time-averaged accuracy loss is degraded, when increasing the value of $\Gamma$, while for a given $N$, the accuracy loss performance is slightly deteriorated.
To understand such a performance, note that the larger is the value of $N$, the smaller is the value of $|{\mathcal C}_t^n|$ and, hence, the smaller is the number of participants, which contributes to the accuracy loss.

Note that our proposed \methodname{} approach emphasizes the control policy determination process and actually joint FL server and client selection algorithm.
Therefore, the actual implementation of this control policy determination can be integrated as a selection component within existing FL platforms.
In the following, to gain insight into the proposed \methodname{} approach for WMSFLN, Figures \ref{fig:4} and \ref{fig:5} compare the performance of \methodname{} with the aforementioned five approaches.
Figures \ref{fig:4a} and \ref{fig:4b} depict the time-averaged WMSFLN cost and accuracy loss in scenario 1, respectively, whereas Figure \ref{fig:5} shows the corresponding results under scenario 2.
As shown in Figure \ref{fig:4b} and Figure \ref{fig:5b}, the performance gap of time-averaged accuracy loss between \methodname{} and remaining five approaches is significant, and it becomes smaller with increasing the value of balance parameter $V$.
However, the reduction of the time-averaged WMSFLN cost for \methodname{} is slight compared to other approaches. Especially as shown in Figure \ref{fig:5a}, it is difficult to distinguish the curves for \methodname{} ($V=0.1, 50$), NCF, and Random.
From the results we notice furthermore that, for both time-averaged WMSFLN cost and accuracy loss, it is observed that \methodname{} for $V=50$ outperforms the remaining approaches.
That is to say, for any given scenario, our proposed \methodname{} has the capability to enhance the performance in terms of time-averaged WMSFLN cost and accuracy loss by choosing the optimal value for the balance parameter $V$.

Finally,  we analyze the performance of the comparison approaches in terms of time-average WMSFLN cost, test accuracy, and fairness. The results are shown in Table \ref{table:1}.
In terms of both test accuracy and fairness, it can be observed that \methodname{} outperforms all the baselines under all experiment conditions.
Specifically, it achieves an average test accuracy and fairness that are 6.91\% and 0.63\% higher, respectively, compared to the best-performing baselines: Fixed delegation and EA.
With respect to the time-average WMSFLN cost, \methodname{} notably outperforms all baseline methods in S1. Furthermore, in S2, \methodname{} consistently achieves top-2 ranks in terms of WMSFLN cost across the MNIST and CIFAR-10 datasets.
On average, it achieved a 27.34\% cost reduction compared to the best-performing baseline, NCF, which prioritizes minimizing cost with no regard to fairness.

\section{Conclusions}
In this paper, we proposed the \methodname{} approach based on Contract Theory and Lyapunov optimization to perform joint task delegation and incentivization in WMSFLN.
To guarantee fair task delegation in WMSFLN, \methodname{} introduced virtual queues to tracked previous access to FL tasks and are updated based on the performance outcomes of the resultant FL models.
The objective is to minimize the time-averaged cost in a WMSFLN, while ensuring all queues remain stable.
This is challenging due to the limited information on the FL clients' participation cost and the volatile nature of WMSFLN states, influenced by the locations of mobile clients.
Extensive experimental results show that \methodname{} achieved 27.34\% lower cost, 6.91\% higher test accuracy, and 0.63\% higher fairness on average than the best-performing baseline.
To the best of our knowledge, it is the first Contract Theory-based FL fair task delegation approach that supports multiple FL servers with limited wireless network coverage coordinating FL training involve FL clients with mobility.


\end{document}